\documentclass[%
 reprint,
 amsmath,amssymb,
 aps,
]{revtex4-1}

\usepackage{graphicx}% Include figure files
\usepackage{dcolumn}% Align table columns on decimal point
\usepackage{bm}% bold math
\usepackage{color}
\newcommand{\be}{\begin{equation}}
\newcommand{\ee}{\end{equation}}
\newcommand{\ba}{\begin{align}}
\newcommand{\ea}{\end{align}}
\newcommand{\ma}{\mathcal}
\newcommand{\MM}{M^2/M_W^2}
\newcommand{\mM}{m^2/M_W^2}

\begin{document}

%\preprint{APS/123-QED}
\title{Contributions of the $\bm{W}$-boson propagator to the $\bm{\mu}$ and $\bm{\tau}$ leptonic decay rates}
% Force line breaks with \\

\author{Andrea Ferroglia}
 \email{aferroglia@citytech.cuny.edu}
\affiliation{%
New York City College of Technology,\\
300 Jay Street, Brooklyn, NY 11201 USA
}%

\author{Christoph Greub}
 \email{greub@itp.unibe.ch}
\affiliation{%
Albert Einstein Center for Fundamental Physics, Institute for Theoretical Physics,\\
University of Bern, CH-3012 Bern, Switzerland
}%

\author{Alberto Sirlin}%
 \email{as6@physics.nyu.edu}
\affiliation{%
 Department of Physics, New York University,\\
 4 Washington Place, New York, NY 10003 USA 
}%

\author{Zhibai Zhang}
 \email{zzhang2@gc.cuny.edu}
\affiliation{%
The Graduate School and University Center, The City University of New York\\ 365 Fifth Avenue, New York NY 10016, USA
}%
\affiliation{%
New York City College of Technology.\\
300 Jay Street, Brooklyn, NY 11201 USA}%

\date{\today}% It is always \today, today,

\begin{abstract}
We derive closed  expressions and useful expansions for the contributions of the tree-level $W$ boson propagator to the the muon and $\tau$ leptonic decay rates. Calling $M$ and $m$ the masses of the initial and final charged leptons, our results in the limit $m=0$ are valid to all orders in $M^2/M_W^2$. In the terms of $\ma O(m_j^2/M_W^2)$ $(m_j=M,m)$, our leading corrections, of $\ma O(M^2/M_W^2)$, agree with the canonical value $(3/5) M^2/M_W^2$, while the coefficient of our subleading contributions, of $\ma O(m^2/M_W^2)$, differs from that reported in the recent literature. A possible explanation of the discrepancy is presented. The numerical effect of the $\ma O(m_j^2/M_W^2)$ corrections is briefly discussed. A general expression, valid for arbitrary values of $M_W$, $M$ and $m$ in the range $M_W>M>m$, is given in the Appendix. 
The paper also contains a review of the traditional definition and evaluation of the Fermi constant.
\end{abstract}

\pacs{Valid PACS appear here}% PACS, the Physics and Astronomy
\maketitle

%\tableofcontents
The correction of $\ma O(m_{\mu}^2/M_W^2)$ to the muon decay rate, arising from the tree-level $W$-boson propagator, is well known in the literature and amounts to a correction factor $1+(3/5)m_{\mu}^2/M_W^2$. An analogous result was first derived by Lee and Yang in the framework of non-local extensions of the Fermi theory \cite{LeeYang}. Calling $M$ and $m$ the masses of the initial and final leptons, recent papers have included both the leading corrections, of $\ma O(M^2/M_W^2)$, as well as the subleading contributions, of $\ma O(m^2/M_W^2)$, to the $\mu$ and $\tau$ leptonic decay rates \cite{KimYew,Krawczyk:2004na,Asner:2008nq}. 

In the present paper, we evaluate the corrections to the $\mu$ and $\tau$ leptonic decay rates induced by the $W$-boson propagator in two cases: {\em i)} in the limit $m=0$, we derive a closed expression, valid to all orders in $M^2/M_W^2$, as well as a useful expansion in powers of $\MM$; {\em ii)} in the corrections of $\ma O(m_j^2/M_W^2)$ $(m_j=M, m)$, we evaluate the leading contributions, of $\ma O(\MM)$, as well as the subleading ones, of $\ma O(\mM)$. In the calculation of the latter, it is important to include the contribution of the $-q^{\mu}q^{\nu}/M_W^2$ term in the unitary-gauge $W$-boson propagator or, equivalently, in other gauges, that of the associated Goldstone boson. In fact, this term leads to contributions of $\ma O(\mM)$. Our result for {\em ii)} is compared with those reported in the recent literature. In the Appendix we present expressions valid for arbitrary values of $M_W$, $M$ and $m$, in the range $M_W>M>m$.  

We focus our attention on $\mu$ decay and later on we extend the results to the $\tau$ leptonic decay rates in a straightforward manner. 
Defining 
\be
x=\frac{M^2}{M_W^2} \, ,  \label{eqn2}
\ee
the terms of ${\mathcal O}(x^n)$ ($n \ge 1$) are very small. For this reason they are evaluated at the tree level, i.~e. to zeroth order in $\alpha$. On the other hand, the  QED correction $\delta_\mu$ to muon decay in the V-A Fermi theory is very important in the term of zeroth order in $x$. In order to obtain simple expressions, we follow the usual procedure of factorizing out the QED correction $[1 + \delta_\mu]$ in all the terms of order $x^n$ ($n \ge 0$) (see, for example \cite{Marciano:1988vm}). This factorization induces terms of ${\mathcal O}(\alpha x^n)$ ($n \ge 1$), which are however extremely small. 
As a consistency check, we have carried out the calculations of the decay rate in two different ways:  the first one is based on a method  described in detail by Veltman in \cite{Veltman}.  
 The method requires to work first in the neutrino pair rest frame, where all scalar products can be written in terms of the energy transferred to the neutrino pair and the angle between the muon and  $\bar{\nu}_e$ momenta. The integral over the $\bar{\nu}_e$  momentum is then carried out using the three-dimensional delta function. 
The angular integrals over the $\nu_\mu$ momentum are trivial, while the integral over its absolute value is implemented by employing the residual one-dimensional $\delta$-function. Finally, one can rewrite the energy transferred to the neutrino pair in a Lorentz invariant way and carry out the integration over the electron momentum in the muon rest frame.
In the second, more conventional approach, one works always in the muon rest frame. We integrate first over the $\bar{\nu}_e$ momentum, thus reducing the four-dimensional $\delta$-function to a one-dimensional one, which leads to the relation $E_2=M(E_0-E)/[M-E+|\vec p|\cos \alpha]$, where $E_2$ is the energy of the muon neutrino, $E$ and $\vec p$ are the energy and momentum of the electron, $E_0=(M^2+m^2)/2M$ its end-point energy, and $\alpha$ the angle between $\vec p$ and the $\nu_{\mu}$ momentum. We then integrate over $E_2$ using the one-dimensional $\delta$-function, over the angle $\alpha$, and finally over $\vec p$. The two approaches lead to the same results, which we present below.
In this paper, we call $\Gamma_{(W)}$ the decay rate when the contributions of the tree-level $W$-boson propagator are included.

{\em i)} Integrating over  the full $W$-boson propagator, in the limit $m \to 0$ we find the closed expression:
\be
\Gamma_{(W)}=\Gamma_0 \left\{\frac{12}{x^3}\left[1-\frac {x}{2}-\frac{x^2}{6}+\frac{\left(1-x\right)}x\ln\left(1-x\right)\right]\right\} \,,\label{eqn1}
\ee
where 
\be
\Gamma_0 = \frac{G_{\mu}^2M^5}{192{\pi}^3} \left[1 + \delta_\mu\right]\, .
\label{Gammazero}
\ee
Furthermore, 
\begin{align}
\frac{G_\mu}{\sqrt{2}} &= \frac{g^2}{8 M_W^2} \left( 1 + \Delta r\right) \, , \label{eqnGmu}
\end{align}
where $g$ is the $\mbox{SU}(2)_L$ gauge coupling constant, $\Delta r$ the electroweak correction introduced in Ref.~\cite{Sirlin:1980nh}, and, as mentioned before, $\delta_\mu$ is the QED correction to muon decay evaluated in the Fermi V-A theory.

Expanding $\ln (1-x)$, Eq.~(\ref{eqn1}) leads to a useful and quickly convergent expression:
\begin{align}
\Gamma_{(W)}&= \Gamma_0 \sum ^{\infty}_{n=0}\frac{12x^n}{(n+3)(n+4)}&\nonumber\\
&= \Gamma_0  \Biggl\{\!1+\!\frac35x+\!\frac25x^2+\!\frac27x^3+\!\frac3{14}x^4
 +\!\frac{x^5}{6}+\cdots \!\Biggr\} \, .&\label{eqn3}
\end{align}
We note that the term of $\ma O(x)$ in Eq.~(\ref{eqn3}) agrees with the canonical result $(3/5) \MM$. Eq.~(\ref{eqn3}) extends that result to all orders in $x$. Since Eq.~(\ref{eqn1}) involves sharp cancellations, the expansion in Eq.~(\ref{eqn3}) is much more useful for numerical calculations. 

{\em ii)} For $m\neq 0$, to zeroth order in $x$, the decay rate is given by the well-known expression
\be
\Gamma_{(W)}^{(0)}= \Gamma_0 F(y)\,,\label{eqn4}
\ee
where 
\be
y=\frac{m^2}{M^2}\,,\label{eqn5}
\ee
and 
\be
F(y)=1-8y-12y^2\ln y+8y^3-y^4\,,\label{eqn6}
\ee
is a phase-space factor (see, for example, Ref. \cite{Sirlin:2012mh}). Eqs.~(\ref{eqn4}-\ref{eqn6}) correspond to the $M_W\rightarrow\infty$ limit and are the usual result in the V-A theory. In order to evaluate the terms of $\ma O(x=\MM)$ with $m\neq 0$, in the calculation we include the correction factor $(1+2q^2/M_W^2)$ arising from the expansion of the $W$-boson propagator, as well as the contribution of the $-q^{\mu}q^{\nu}/M_W^2$ term in the propagator. This leads to the simple and compact result:
\be
\Gamma_{\left(W\right)}^{\left(1\right)}= \Gamma_0 \frac35x(1-y)^5\,.\label{eqn7}
\ee
An interesting theoretical feature of Eq.~(\ref{eqn7}) is that logarithmic terms proportional to $\ln y$ cancel between the contributions of the $(1+2q^2/M_W^2)$ correction factor and the $-q^{\mu}q^{\nu}/M_W^2$ term in the propagator. We also observe that the $y$ dependence in Eq.~(\ref{eqn7}) is very different from that in Eqs.~(\ref{eqn4},\ref{eqn6}), so that in their sum $F(y)$ does not factorize in a simple manner. Neglecting terms of $\ma O\left(xy^2=m^4/(M_W^2M^2)\right)$ and higher, Eq.~(\ref{eqn7}) reduces to 
\be
\Gamma_{(W)}^{(1)}=  \Gamma_0 \left\{\frac35\frac{M^2}{M_W^2}-\frac{3m^2}{M_W^2}\right\}\,.\label{eqn8}
\ee

The leading correction, $(3/5) \MM$, agrees once more with the canonical result. In the subleading correction of $\ma O(m^2/M_W^2)$, $-2m^2/M_W^2$ arises from the contribution of the $-q^{\mu}q^{\nu}/M_W^2$ term, while an additional $-m^2/M_W^2$ is induced by the $(1+2q^2/M_W^2)$ correction factor. 

In the muon decay case, $M$ and $m$ are identified with $m_{\mu}$ and $m_e$. The extension of our results to the $\tau$ leptonic decay rates is straightforward: $M$ is identified with $m_{\tau}$, while $m=m_{\mu}$ in $\Gamma_{(W)}(\tau\rightarrow\nu_{\tau}+\mu+\bar{\nu}_{\mu})$ and
$m=m_e$ in $\Gamma_{(W)}(\tau\rightarrow\nu_{\tau}+e+\bar{\nu}_e)$. Furthermore,  $\delta_\mu$ should be changed into $\delta_\tau$, namely the  appropriate QED corrections in $\tau$ decays.

As far as we know, our calculation {\em i)}, valid to all orders in $\MM$ in the $m\rightarrow 0$ limit, has not been carried out in the literature. In order to compare our calculation {\em ii)} with existing results, we combine Eqs.~(\ref{eqn4}, \ref{eqn8}): 
\begin{align}
\Gamma^{(0)}_{(W)}+\Gamma^{(1)}_{(W)}&= \Gamma_0 \Biggl[F(y)+\frac35\frac{M^2}{M_W^2}-\frac{3m^2}{M_W^2}&\nonumber\\
&+\ma O \left(\frac{m^4}{M_W^2M^2} \right)\Biggr]\,.&\label{eqn9}
\end{align}
In the literature, the phase space factor $F(y)$ is often factorized. Performing such factorization, Eq.~(\ref{eqn9}) becomes  
\begin{align}
\Gamma^{(0)}_{(W)}+\Gamma^{(1)}_{(W)}&= \Gamma_0 F(y)\Biggl[1+\frac35\frac{M^2}{M_W^2}+\frac95\frac{m^2}{M_W^2}&\nonumber\\
&+\ma O\left(\frac{m^4}{M_W^2M^2}\right)\Biggr]\,.&\label{eqn10}
\end{align}
Thus, the factorization of $F(y)$ induces a large change in the coefficient of the subleading correction of $\ma O(\mM)$. This is easy to understand recalling Eqs.~(\ref{eqn5}, \ref{eqn6}): through terms of $\ma O(m^2/M^2)$ the factorization of $F(y)$ effectively leads to the change  
\begin{align}
\frac35\frac{M^2}{M_W^2}\rightarrow\frac35\frac{M^2}{M_W^2}\left(1+\frac{8m^2}{M^2}\right)=\frac35\frac{M^2}{M_W^2}+\frac{24}5\frac{m^2}{M_W^2}\,.\label{eqn11}
\end{align}
As a consequence, the factorization of $F(y)$ induces a new subleading correction $\left(24/5\right) m^2/M_W^2$ in the expression between square brackets in Eq.~(\ref{eqn10}). Combining this with $-3m^2/M_W^2$ in Eq.~(\ref{eqn9}), one obtains the subleading correction $(9/5) m^2/M_W^2$ reported in Eq.~(\ref{eqn10}).

Our result in Eq.~(\ref{eqn10}) can be compared with expressions published in the recent literature. For example, Refs. \cite{KimYew,Krawczyk:2004na,Asner:2008nq} consider the decay $\tau\rightarrow l+\bar{\nu}_l+\nu_{\tau}$ $(l=\mu,e)$. Modulo QED corrections, the result for the leptonic decay rates presented in those papers  is: 
\be
\Gamma^l=\frac{G_{\mu}^2m_{\tau}^5}{192\pi^3} f\left(\frac{m_l^2}{m_{\tau}^2}\right)\left(1+\frac35 \frac{m_{\tau}^2}{M_W^2}-\frac{2m_l^2}{M_W^2}\right)\,,\label{eqn12}
\ee
while, in this case, our Eq.~(\ref{eqn10}) becomes
\begin{align}
\Gamma^{(0) l}_{(W)}+\Gamma^{(1) l}_{(W)}&=\frac{G_{\mu}^2m_{\tau}^5}{192\pi^3}F\left(\frac{m_l^2}{m_{\tau}^2}\right)\Biggl[1+\frac35\frac{m_{\tau}^2}{M_W^2}+\frac95\frac{m_l^2}{M_W^2}&\nonumber\\
&+\ma O\left(\frac{m_l^4}{M_W^2m_{\tau}^2}\right)\biggr]\,.&\label{eqn13}
\end{align}
Since the function $f$ in Eq.~(\ref{eqn12}) is identical to $F$,
% i.e. $f(m_l^2/m_{\tau}^2)=F(m_l^2/m_{\tau}^2)$, 
we see that the two results agree on the leading correction $(3/5) m_{\tau}^2/M_W^2$, but sharply disagree on the coefficient of the subleading term of $\ma O(m_l^2/M_W^2)$. In particular, the signs of the $\ma O(m_l^2/M_W^2)$ correction are opposite.

A possible explanation of this difference could be that: {\em 1)} in the derivation of Eq.~(\ref{eqn12}), only the subleading $-2m_l^2/M_W^2$ contribution from the $-q^{\mu}q^{\nu}/M_W^2$ term in the propagator has been retained (thus neglecting the additional $- m_l^2/M_W^2$ contribution arising from the $(1+ 2 q^2/M_W^2)$ correction factor) and {\em 2)} the additional $(24/5) m_l^2/M_W^2$ contribution induced by the factorization of $F(y)$ has not been taken into account. 

Numerically, the corrections of $\ma O(m_j^2/M_W^2)$ $(m_j=M, m)$ are very small. Their largest values are attained in the decay $\tau\rightarrow\mu+\bar{\nu}_{\mu}+\nu_{\tau}$. In this case, the correction factor is 
\be
1+\frac 35\frac{m_{\tau}^2}{M_W^2}+\frac 95\frac{m_{\mu}^2}{M_W^2}=1+2.9315\times 10^{-4}+3.11\times10^{-6}\,,\label{eqn14}
\ee
where we employed $M_W=80.385~\mbox{GeV}$. Since the current relative error in the measurement of the $\tau$ lifetime is $3.44\times 10^{-3}$, in order to be sensitive to the leading correction in Eq.~(\ref{eqn14}), it would be necessary to decrease the experimental error by more than a factor $10$.

In the case of muon decay
\be
1+\frac 35\frac{m_{\mu}^2}{M_W^2}+\frac 95\frac{m_{e}^2}{M_W^2}=1+1.0366\times 10^{-6}+7.3\times 10^{-11}\, .\label{eqn15}
\ee
The current relative error in the measured muon lifetime is $1.00\times 10^{-6}$ \cite{Mulan}. Thus, the leading correction in Eq.~(\ref{eqn15}) is very close to the experimental error; it is also very close to the two-loop QED correction (see, for example, Eq.~(36) in Ref. \cite{Sirlin:2012mh}). Thus, at present, the $\ma O(m_{\mu}^2/M_W^2)$ correction has a marginal effect in muon decay. On the other hand, in the foreseeable future the subleading corrections of $\ma O(\mM)$ are out of experimental reach in both $\mu$ and $\tau$ decays.

We remind the reader that,  in the traditional approach, the Fermi constant $G_F$ is defined from the muon lifetime, as evaluated in the Fermi V-A theory  to first order in the weak interaction coupling constant.
Specifically, $G_F$ is defined by the relation
\begin{align}
\frac{1}{\tau_\mu} &= \frac{G_F^2 m_\mu^5}{192 \pi^3} F(y) \left[ 1+ \delta_\mu \right] \, ,
\label{eqnlife}
\end{align}
where $\tau_\mu$ is the muon lifetime, and $\delta_\mu$ the QED correction.
 This approach has several important advantages (see, for example \cite{Sirlin:2012mh}): {\emph{i)}} the muon lifetime has been measured with great accuracy, {\emph{ii)}} to first order in $G_F$, but all orders in $\alpha$, the very important QED correction to muon decay in the Fermi V-A theory is known to be finite after charge and mass renormalization \cite{BermanSirlin}, \emph{iii)} at present, its contribution to the muon lifetime has been evaluated through two loop order \cite{vanRitbergen:1998yd, vanRitbergen:1999fi, Steinhauser:1999bx} and estimated at three loops \cite{Ferroglia:1999tg},
\emph{iv)} very importantly, in the traditional definition, $G_F$ is a true constant of nature, like the electric charge: it does not need to be redefined and numerically changed every time a new particle contributing to muon decay is discovered, 
\emph{v)} the relation of $G_F$ to the fundamental constants of the Standard Theory of particle physics involves the electroweak radiative correction $\Delta r$ and has been explained in Ref.~\cite{Sirlin:1980nh}.
A detailed description of the current evaluation of $G_F$ is provided in Section II-D of Ref.~\cite{Sirlin:2012mh}. It includes one and two-loop QED corrections treated in two alternative ways, very small contributions of ${\mathcal O}(\alpha)$ and 
${\mathcal O}(\alpha^2)$ proportional to powers of $y$, and an estimate of the theoretical error due to truncation of the QED perturbative series. The current value is~\cite{Mulan}:
\begin{align}
G_F = 1.1663788(7) \times 10^{-5}~\text{GeV} \, ,
\end{align}
an important determination at the $0.6$~ppm level.

It is then clear that, in the traditional approach, the corrections
from the $W$-boson propagator we discuss in this paper do not affect the definition or the value of $G_F$. Rather, they are interpreted as additional, albeit very small, corrections to the $\mu$ and $\tau$ leptonic decay rates that emerge in the Standard Theory of particle physics.
In fact, writing $\Gamma_{(W)}$ in the form
\begin{equation}
\Gamma_{(W)} \equiv \frac{G_\mu^2 m_\mu^5}{192 \pi^3}  F(y) \left[1 + \delta_\mu\right] \left(1+ \delta_{(W)} \right) \, ,
\label{eqn24}
\end{equation}
and comparing Eqs.~(\ref{eqnlife}, \ref{eqn24}), one finds that 
the relation between $G_F^2$ and $G_\mu^2$ is given by
\begin{align}
G_F^2 &= G_\mu^2 \left(1+ \delta_{(W)} \right) \, , \nonumber \\
&= \frac{g^4}{32 M_W^4} \left(1+ \Delta r \right)^2  \left( 1+ \delta_{(W)} \right) \, , \label{eqn20}
\end{align}
where, in the last step,  we employed Eq.~(\ref{eqnGmu}). The last factor in Eqs.~(\ref{eqn24}, \ref{eqn20}) represents the additional tree-level correction induced by the $W$-boson propagator in the Standard Theory.

In summary, we have re-examined the contribution of the $W$-boson
propagator to the $\mu$ and $\tau$ leptonic decay rates. Calling $M$
and $m$ the masses of the initial and final charged leptons, in the
limit $m\rightarrow 0$ we have derived a closed expression,
Eq.~(\ref{eqn1}), and a useful expansion, Eq.~(\ref{eqn3}), valid to
all orders in $x=\MM$. They extend the canonical result $(3/5) M^2/
M_W^2$  to all orders in $x$. In the terms of $\ma O(m_j^2/M_W^2)$
$(m_j=M, m)$, we have evaluated the leading corrections, of $\ma
O(\MM)$, as well as the subleading ones, of $\ma O(\mM)$
(Eq.~(\ref{eqn10})). While our leading corrections agree with the
canonical result, the coefficient of our subleading corrections
differs sharply from the one reported in the recent literature. A
possible explanation of this discrepancy was presented. The numerical
effect of the $\ma O(m_j^2/M_W^2)$ corrections was briefly
discussed. In the Appendix we have presented an expression for the 
leptonic decay rates that includes
the contribution of the $W$-boson propagator for arbitrary values of $M_W$, $M$ and $m$ in the range $M_W>M>m$ (Eq.~(\ref{eqn18})). The paper also contains a review of the traditional definition and evaluation of the Fermi constant.

We would like to thank M.~Passera for calling our attention to reference~\cite{KimYew} and for useful discussions.
The work of A.~F. and Z.~Z. was supported in part
by the National Science Foundation Grant No. PHY-1068317.
The work of A. S. was supported in part by the National
Science Foundation Grant No. PHY-0758032. The work of C.~G. was
partially supported by the Swiss National Science Foundation.

%%%%%%%%%%%%%%%%%%%%%%%%%%%%%%%%%%%%%%%%%%%%%%
\appendix

\section*{Appendix}

In this Appendix we present expressions for $\Gamma_{(W)}$, valid for arbitrary values of $M_W$, $M$ and $m$, in the range $M_W>M>m$. We find 
\begin{widetext}
\begin{align}
\Gamma_{(W)} &= \Gamma_0 \,  3 \, \Biggl\{  \frac{4(1-y)}{x^3}
          - \frac{2 (1-y^2)}{x^2}
           - \frac{1}{x} \Biggl(\frac{2}{3}
          - y
          + y^2
%\nonumber \\ &
          - \frac{2 }{3}y^3 \Biggr)\!
          - \frac{5 }{2}y (1\!-\!y^2)
          + x y^2 (1\!-\!y)
          - y^2 \left(4 \!+\! x^2 y\right)  \ln y 
          \nonumber \\ &+\frac{1}{x^4}\ln\left(\frac{1-x}{1 -x y} \right) \Bigl[ 
                    4(1- x) - x  y \bigl(4- 3 x
- x^3(1 
%\nonumber \\ &
+ y^2(1- x^2)) \bigr)\Bigr] \Biggr\} \, , \label{eq:full}
\end{align}
\end{widetext}
where $\Gamma_0$ is given in Eq.~(\ref{Gammazero}). 
By setting $y=0$ in Eq.~(\ref{eq:full}),  one immediately  finds the result in Eq.~(\ref{eqn1}). Eq.~(\ref{eqn4}) can be recovered 
by taking the $x \to 0$ limit in Eq.~(\ref{eq:full}).
Although the decay rate is obviously well behaved in the limit $x \to 0$, several of the terms in Eq.~(\ref{eq:full}) are singular in that limit. This fact gives rise to large cancellations among different terms, which in turn lead to a loss of significant digits in numerical evaluations. This problem can be avoided by rewriting the second logarithm in Eq.~(\ref{eq:full}) as an infinite sum
\be
\ln\left(\frac{1-x}{1 -x y} \right) = - \sum_{n=1}^{\infty}  (1-y^n) \frac{x^n}{n} \, .
\ee
The first three terms in this series, when inserted in Eq.~(\ref{eq:full}), lead to the cancellation of all terms which are singular in the $x \to 0$ limit. After a few manipulations of the residual series, $\Gamma_{(W)}$ can be written in the form
%
%\begin{widetext}
\begin{align}
\Gamma_{(W)} &=\Gamma_0   \Biggl\{ F(y) + \frac{3}{5} x (1-y)^5  + \frac{x^2}{20} \Bigl( 8 -27 y 
\nonumber \\ &
+ 27 y^5
 -8 y^6 - 60 y^3 \ln{y} \Bigr)
 +3 \sum_{n=3}^\infty x^n H_n\left(y\right) \Biggl\}  \, , \label{eqn18}
\end{align}
%\end{widetext}
where $F(y)$ is defined in Eq.~(\ref{eqn6}) and 
%
%\begin{widetext}
\begin{align}
H_n\left( y\right) &=\frac{y^3 (1-y^{n-2})}{n-2}  
 - \frac{y (1+y^2) (1-y^n)}{n} 
 \nonumber \\ & - \frac{3 y (1-y^{n+2}) }{n+2}
 + \frac{4 (1+y) (1-y^{n+3})}{n+3}
 \nonumber \\ &  
 - \frac{4  (1-y^{n+4})}{n+4} \, .
\end{align}
%\end{widetext}
%
The contributions of ${\mathcal O}(1,x,x^2)$ are shown explicitly in the first three terms of Eq.~(\ref{eqn18}),
while those of ${\mathcal O}(x^n)$ ($n \ge 3$) are given in the series presented at the end of the equation.
An interesting property of the functions $H_n(y)$ ($n \ge 3$) is that they are proportional to $(1-y)^5$. This is due to the fact that $H_n(y)$ and its first four derivatives vanish at $y=1$. The same property holds for the contribution of ${\mathcal O}(x)$, as explicitly shown in the second term of  Eq.~(\ref{eqn18}). It is also interesting to observe that only the contributions of ${\mathcal O}(x^0)$ and of ${\mathcal O}(x^2)$ contain terms proportional to $\ln y$.

% so that the decay rate can be rewritten as
%
%OLD VERSION - to be delated
%
%\begin{align}
%\Gamma_{(W)} &=\frac{G_{\mu}^2 M^5}{192 \pi^3} \Biggl\{4-x^5 y^6+x^5 y^3-\frac{3 x^4 y^5}{2}+\frac{3 x^4 y^3}{2}
%\nonumber \\ &+x^3 y^6-2
%   x^3 y^4+2 x^3 y^3-x^3 y+\frac{3 x^2 y^5}{2}-\frac{3 x^2 y}{2}
%   \nonumber \\ &+6 x
%   y^4-6 x y^3+6 x y^2-6 x y-4 y^4+8 y^3-8 y 
%\nonumber \\ &- 3 y^2 \ln y \left[ 4 + x^2 y\right]  - 3 \sum_{n=4}^{\infty}  (1-y^n) \frac{x^{n-4}}{n} \times 
%\nonumber \\ &\times \Bigl[ 
%                    4- 4 x - 4x y+ 3 x^2y+ x^4 y + x^4 y^3- x^6 y^3\Bigr]
%\Biggl\} \, .
%\end{align}

%%%%%%%%%%%%%%%%%%%%%%%%%%%%%%%%%%%%%%%%%%%%%%

\end{document}